\documentstyle[aps]{revtex}

\begin{document}

\noindent {\bf RADIATION FROM PERFECT MIRRORS FOLLOWING PRESCRIBED
RELATIVISTIC TRAJECTORIES (*)}

\bigskip

\noindent {\it A. Calogeracos}

\noindent {\it Division of Theoretical Mechanics}

\noindent {\it Hellenic Air Force Academy TG1010}

\noindent {\it Dhekelia Air Force Base}

\noindent {\it Dhekelia, Greece}

\medskip

\noindent (*)\ Talk presented at the Fifth Workshop on ''{\it Quantum Field
Theory under the Influence of External Conditions'',} Leipzig, September
2001, published in Int. J. of Mod. Phys. {\bf A17}, 1018 (2002)

\medskip

\noindent {\bf Abstract}

\smallskip

The question is examined of a mirror which starts from rest and either (i)
accelerates for some time and eventually reverts to motion at constant
velocity, (ii) continues accelerating forever. A sharp distinction is made
between cases (i) and (ii) concerning the spectrum of the emitted radiation,
and the qualitative difference between the two cases is pointed out. The
Bogolubov coefficients are calculated for a trajectory of type (i). A type
(ii) trajectory is entirely unphysical as far as any realistic mirror is
concerned, however it is of interest in that it has been used as a simple
analog of black hole collapse. The spectrum emitted for the type (ii)
trajectory $z=-\ln \left( \cosh t\right) $ is examined and it is shown that
it is indeed that of a black body. Inconsistencies in previous derivations
of the above result are pointed out.

\section{Introduction}

The subject of mirror induced radiation has received great attention in the
last ten years. Nonrelativistic calculations have been performed for
imperfect mirrors (dielectric or dispersive) using the Hamiltonian
formalism. In this talk, rather than report on recent progress along these
lines, I shall go about twenty five years back and examine the old problem
of an one-sided perfect mirror following a prescribed relativistic
trajectory. The papers by Fulling and Davies \cite{FD}, \cite{DF} are early
important contributions on the subject, and a good review is provided by
Birrell and Davies \cite{BD}. I consider a mirror that starts from rest and
accelerates along the trajectory $z=-\ln \left( \cosh \kappa t\right) $ (I
take $\kappa =1$ thus fixing the energy scale). I make the distinction
between a type (i) (asymptotically inertial) trajectory where the mirror
accelerates till it reaches a space-time point {\it P }and then reverts to
motion at uniform velocity $B_{P}$ and a type (ii) trajectory where the
mirror accelerates ad infinitum. The results presented here are contained in
the two preprints \cite{AC1}, \cite{AC2}.

Concerning the type (i) case an exact calculation of the Bogolubov $\beta
\left( \omega ,\omega ^{\prime }\right) $ amplitude in terms of
hypergeometric functions is presented, the analytic properties of the $%
\alpha \left( \omega ,\omega ^{\prime }\right) $ and $\beta \left( \omega
,\omega ^{\prime }\right) $ amplitudes are discussed and the constraints
imposed by unitarity are stressed. There are several advantages in
considering asymptotically inertial trajectories: (a) Acceleration
continuing for an infinite time implies mathematical singularities and also
entails physical pathologies associated, for example, with the infinite
energy that has to be imparted to the mirror. (b) The mirror's rest frame
eventually (after acceleration stops) becomes an inertial frame and the
standard description in terms of {\it in }and {\it out }states is possible.
One may choose either the lab frame or the mirror's rest frame to describe
the photons produced. (c) One avoids statements about photons produced {\it %
while }the mirror is accelerated; rather one makes unambiguous statements
pertaining to times $t=\pm \infty $ when acceleration vanishes. We show in
general that for an asymptotically inertial trajectory with the velocity
being everywhere continuous the amplitude squared $\left| \beta \left(
\omega ,\omega ^{\prime }\right) \right| ^{2}$ goes as $\left( \omega
^{\prime }\right) ^{-5}$ (or faster) for large $\omega ^{\prime }.$ This is
in contrast to a type (ii) trajectory where $\left| \beta \left( \omega
,\omega ^{\prime }\right) \right| ^{2}$ goes as $1/\omega ^{\prime }$ for
large $\omega ^{\prime }$. This radical difference is due to a subtle
cancellation in the type (i) case between two contributions, one arising
from the initial and the other from the final asymptotic part of the
trajectory.

The type (ii) trajectory mentioned previously has been considered
extensively in the past, starting with the classic papers by Fulling and
Davies op. cit.. The problem is of interest because of its connection to
radiation emitted from a collapsing black hole (Hawking \cite{HAW}, DeWitt 
\cite{dew}) and to the attendant thermal spectrum. In the present context
one can counter that such a trajectory is unrealistic on the grounds of (b)
above; then one may decide to stick to type (i) trajectories. In the second
part of this talk I ignore such questions, take the premises of the early
papers on the subject for granted, and concentrate on the calculation of the
Bogolubov amplitude $\beta (\omega ,\omega ^{\prime })$ (and thus of $\alpha
(\omega ,\omega ^{\prime })$ as well) and of the spectrum 
\begin{equation}
N(\omega )=\int_{0}^{\infty }d\omega ^{\prime }\left| \beta (\omega ,\omega
^{\prime })\right| ^{2}e^{-a\left( \omega +\omega ^{\prime }\right) }
\label{bb2}
\end{equation}

\noindent ($a$ is a convergence factor). We shall show that contrary to
folklore a correct derivation of the black body spectrum via the calculation
of the Bogolubov amplitudes requires consideration of the whole trajectory
and not just of its asymptotic part.

\section{The Bogolubov amplitudes for a type (i) trajectory}

\subsection{Calculation of $\beta \left( \omega ,\omega ^{\prime }\right) $}

Let us introduce coordinates 
\begin{equation}
u=t-z,v=t+z  \label{e01}
\end{equation}

\noindent The point {\it P }at which the mirror reverts to uniform motion
has $v$ coordinate given by $v=r$. The velocity at {\it P} is given by $%
B_{P}=1-e^{r}$ (see the Appendix of \cite{AC1} for details). The trajectory
equation $u=f(v)$ is defined in a piecewise manner via 
\begin{equation}
u=v,v<0  \label{lei1}
\end{equation}

\noindent corresponding to a mirror at rest for $t<0$, 
\begin{equation}
u=f_{acc}(v)\equiv -\ln \left( 2-e^{v}\right) ,0<v<r  \label{lei2}
\end{equation}

\noindent corresponding to the accelerating part of the trajectory, and 
\begin{equation}
u=f_{0}(v)\equiv C+\frac{1-B_{P}}{1+B_{P}}v  \label{lei3}
\end{equation}

\noindent describing uniform motion after point {\it P. }The constant $C$ is
fixed by the requirement that the trajectory be continuous at {\it P} 
\begin{equation}
C+\frac{1-B_{P}}{1+B_{P}}r=-\ln \left( 2-e^{r}\right)  \label{40b}
\end{equation}

\noindent Let us take everything to exist to the right of the mirror. There
are two sets of modes 
\begin{equation}
\varphi _{\omega }(u,v)=%
{\displaystyle {i \over 2\sqrt{\pi \omega }}}%
\left( \exp (-i\omega v)-\exp \left( -i\omega p(u)\right) \right)  \label{e3}
\end{equation}

\noindent and 
\begin{equation}
\bar{\varphi}_{\omega }(u,v)=%
{\displaystyle {i \over 2\sqrt{\pi \omega }}}%
\left( \exp \left( -i\omega f(v)\right) -\exp \left( -i\omega u\right)
\right)  \label{e5}
\end{equation}

\noindent which satisfy the free wave equation and the condition that the
field should vanish on the mirror. The modes $\varphi _{\omega }(u,v)$ of (%
\ref{e3}) describe waves incident from the right as it is clear from the
sign of the exponential in the first term; the second term represents the
reflected part which has a rather complicated behaviour depending on the
motion of the mirror. These modes constitute the {\it in }space and should
obviously be unoccupied before acceleration starts, 
\[
a_{i}\left| 0in\right\rangle =0 
\]
Similarly the modes $\bar{\varphi}_{\omega }(u,v)$ describe waves travelling
to the right (emitted by the mirror) as can be seen from the exponential of
the second term. Correspondingly the first term is complicated. These modes
define the {\it out }space and 
\[
\bar{a}_{i}\left| 0out\right\rangle =0 
\]
The state \noindent $\left| 0out\right\rangle $ corresponds to the state
where nothing is produced by the mirror. The two representations are
connected by the Bogolubov transformation 
\begin{equation}
\bar{a}_{i}=\sum_{j}\left( \alpha _{ji}a_{j}+\beta _{ji}^{*}a_{j}^{\dagger
}\right)  \label{e011}
\end{equation}

\noindent The fact that the {\it in }and {\it out }vacua are not identical
lies at the origin of particle production. In our notation the matrix $\beta
_{ji}$ is given by the overlap (see Birrell and \ Davies op. cit. equations
(2.9), (3.36)) 
\begin{equation}
\beta (\omega ,\omega ^{\prime })=-i\int_{0}^{\infty }dz\varphi _{\omega
^{\prime }}(z,0)%
{\displaystyle {\partial  \over \partial t}}%
\bar{\varphi}_{\omega }(z,0)+i\int_{0}^{\infty }dz\left( 
{\displaystyle {\partial  \over \partial t}}%
\varphi _{\omega ^{\prime }}(z,0)\right) \bar{\varphi}_{\omega }(z,0)
\label{e11}
\end{equation}

\noindent The integration in (\ref{e11}) can be over any spacelike
hypersurface and the choice $t=0$ is convenient. After we substitute
expressions (\ref{e3}), (\ref{e5}) in (\ref{e11}) we get the $\beta (\omega
,\omega ^{\prime })$ amplitude in the form

\begin{eqnarray}
\beta (\omega ,\omega ^{\prime }) &=&%
{\displaystyle {1 \over 4\pi \sqrt{\omega \omega ^{\prime }}}}%
\int_{0}^{\infty }dz\left\{ e^{i\omega ^{\prime }z}-e^{-i\omega ^{\prime
}z}\right\} \left\{ \omega ^{\prime }e^{-i\omega f}-\omega f^{\prime
}e^{-i\omega f}\right\} +  \nonumber \\
&&  \label{bb9} \\
&&+%
{\displaystyle {\left( \omega -\omega ^{\prime }\right)  \over 4\pi \sqrt{\omega \omega ^{\prime }}}}%
\int_{0}^{\infty }dz\left\{ e^{i\omega ^{\prime }z}-e^{-i\omega ^{\prime
}z}\right\} e^{i\omega z}  \nonumber
\end{eqnarray}

\noindent Notice that the second integral is $f$-independent (i.e.
trajectory independent) and that its origin is purely kinematic. Some of the
integrals involved are conveniently expressed in terms of the $\zeta $
function and its complex conjugate $\zeta ^{*}$defined in Heitler \cite{HE},
pages 66-71: 
\begin{equation}
\zeta (x)\equiv -i\int_{0}^{\infty }e^{i\kappa x}d\kappa =P\frac{1}{x}-i\pi
\delta (x)  \label{delta}
\end{equation}

Amplitude (\ref{bb9}) is split to three contributions

\begin{equation}
\beta (\omega ,\omega ^{\prime })\equiv \beta _{I}(\omega ,\omega ^{\prime
})+\beta _{II}(\omega ,\omega ^{\prime })+\beta _{III}(\omega ,\omega
^{\prime })  \label{e50}
\end{equation}

\noindent originating as follows. The $\beta _{III}(\omega ,\omega ^{\prime
})$ term stands for the second ($f$-independent) integral in (\ref{bb9}), is
always there for a mirror starting from rest (or initially moving at uniform
velocity), and does not depend on the subsequent form of the trajectory. The 
$\beta _{I}(\omega ,\omega ^{\prime })$ contribution results from the
accelerating part of the trajectory and corresponds to the $0<z<r$
integration range in (\ref{bb9}). Its evaluation is mathematically somewhat
involved. The $\beta _{II}(\omega ,\omega ^{\prime })$ amplitude results
from the $z>r$ part of the integration range in (\ref{bb9}), is readily
evaluated, and is specific to a mirror that reverts to the state of uniform
motion (it does not arise in the case of a type (ii) trajectory). Thus

\begin{equation}
\beta _{I}(\omega ,\omega ^{\prime })=-%
{\displaystyle {i \over 2\pi \sqrt{\omega ^{\prime }\omega }}}%
\int_{0}^{r}dz\sin (\omega ^{\prime }z)\left\{ \omega f_{acc}^{\prime
}(z)-\omega ^{\prime }\right\} e^{-i\omega f_{acc}(z)}  \label{e20}
\end{equation}

\[
\beta _{II}(\omega ,\omega ^{\prime })= 
\]

\[
=\frac{1}{4\pi i\sqrt{\omega \omega ^{\prime }}}\left( \omega 
{\displaystyle {1-B_{P} \over 1+B_{P}}}%
-\omega ^{\prime }\right) \exp \left[ i\left( \omega 
{\displaystyle {1-B_{P} \over 1+B_{P}}}%
+\omega ^{\prime }\right) r\right] \zeta \left( \omega 
{\displaystyle {1-B_{P} \over 1+B_{P}}}%
+\omega ^{\prime }\right) e^{-i\omega C}- 
\]

\begin{equation}
-\frac{1}{4\pi i\sqrt{\omega \omega ^{\prime }}}\exp \left[ i\left( -\omega
^{\prime }+\omega 
{\displaystyle {1-B_{P} \over 1+B_{P}}}%
\right) r\right] e^{-i\omega C}  \label{e102}
\end{equation}

\begin{equation}
\beta _{III}(\omega ,\omega ^{\prime })=\frac{1}{4\pi i\sqrt{\omega \omega
^{\prime }}}-\frac{1}{4\pi i\sqrt{\omega \omega ^{\prime }}}\left( \omega
-\omega ^{\prime }\right) \zeta \left( \omega +\omega ^{\prime }\right)
\label{e103}
\end{equation}

\noindent Notice that the arguments of the $\zeta $ functions in (\ref{e102}%
) and (\ref{e103}) never vanish. Hence as far as the evaluation of $\beta
(\omega ,\omega ^{\prime })$ is concerned we observe that it is only the
first term in (\ref{delta}) that is operative, and that $\beta (\omega
,\omega ^{\prime })$ does not have any $\delta $-type singularities. Giving
the result in the form (\ref{e102}), (\ref{e103}) is useful because we can
obtain the other Bogolubov amplitude $\alpha (\omega ,\omega ^{\prime })$
via the substitution $\omega \rightarrow -\omega $ in accordance with (\ref
{e107}) below. The $\alpha (\omega ,\omega ^{\prime })$ is of course
expected to have $\delta $-type singularities, and in fact reduces to just $%
\delta (\omega -\omega ^{\prime })$ in the trivial case when the {\it in }%
and {\it out }modes coincide. Comparing with Davies and Fulling \cite{DF}
note that the $\beta _{III}$ term is unaccountably missing from the latter
reference. It will be shown in the following section that this term is in
fact crucial in containing the thermal spectrum for the type (ii) trajectory.

The amplitude $\beta _{I}$ may be expressed in terms of hypergeometrics (for
details on the calculation see \cite{AC1}) 
\begin{equation}
\beta _{I}(\omega ,\omega ^{\prime })=\frac{1}{2\pi \sqrt{\omega \omega
^{\prime }}}\sin \left( \omega ^{\prime }r\right) \left( 2-e^{r}\right)
^{i\omega }-  \label{e25}
\end{equation}

\[
-\frac{1}{2\pi i}\frac{2^{i\omega }}{\sqrt{\omega \omega ^{\prime }}}%
\{F\left( -i\omega ,-i\omega ^{\prime };-i\omega ^{\prime }+1;\frac{1}{2}%
\right) + 
\]

\[
+F\left( -i\omega ,-i\omega ^{\prime };-i\omega ^{\prime }+1;\frac{e^{r}}{2}%
\right) e^{-i\omega ^{\prime }r}\} 
\]
It is interesting to observe that the value $r=\ln 2$ corresponding to the
asymptote of the type (ii) trajectory (see (\ref{lei2})) falls exactly on
the radius of convergence of the series as one can see from the argument of
the second hypergeometric appearing above.

\subsection{On the Bogolubov coefficients $\alpha \left( \omega ,\omega
^{\prime }\right) ,$ $\beta \left( \omega ,\omega ^{\prime }\right) $}

The Bogolubov $\alpha $ coefficients appearing in (\ref{e011}) are given by

\begin{equation}
\alpha (\omega ,\omega ^{\prime })=i\int_{0}^{\infty }dz\varphi _{\omega
^{\prime }}(z,0)%
{\displaystyle {\partial  \over \partial t}}%
\bar{\varphi}_{\omega }^{*}(z,0)-i\int_{0}^{\infty }dz\left( 
{\displaystyle {\partial  \over \partial t}}%
\varphi _{\omega ^{\prime }}(z,0)\right) \bar{\varphi}_{\omega }^{*}(z,0)
\label{91}
\end{equation}

\noindent Recall also the unitarity condition ((3.39) of Birrell and \
Davies op. cit.) 
\begin{equation}
\int_{0}^{\infty }d\tilde{\omega}\left( \alpha \left( \tilde{\omega},\omega
_{1}\right) \alpha ^{*}\left( \tilde{\omega},\omega _{2}\right) -\beta
\left( \tilde{\omega},\omega _{1}\right) \beta ^{*}\left( \tilde{\omega}%
,\omega _{2}\right) \right) =\delta \left( \omega _{1}-\omega _{2}\right)
\label{e92a}
\end{equation}

\noindent and its partner 
\begin{equation}
\int_{0}^{\infty }d\widetilde{\omega }\left( \alpha \left( \omega _{1},%
\widetilde{\omega }\right) \alpha ^{*}\left( \omega _{2},\widetilde{\omega }%
\right) -\beta \left( \omega _{1},\widetilde{\omega }\right) \beta
^{*}\left( \omega _{2},\widetilde{\omega }\right) \right) =\delta \left(
\omega _{1}-\omega _{2}\right)  \label{92}
\end{equation}
Relations (\ref{e92a}), (\ref{92}) above are direct consequences of the fact
that the sets $\bar{\varphi}_{\omega }$ and $\varphi _{\omega ^{\prime }}$
respectively are orthonormal and complete. They also guarantee that the
operators $a_{j},a_{j}^{\dagger }$ and $\bar{a}_{i},\bar{a}_{i}^{\dagger }$
obey the standard equal time commutation relations that creation and
annihilation operators do.

We find it convenient to isolate the square roots in (\ref{bb9}) and
introduce quantities $A(\omega ,\omega ^{\prime }),B(\omega ,\omega ^{\prime
})$ that are analytic functions of the frequencies (without the branch cuts
attendant to square roots) via 
\begin{equation}
\alpha (\omega ,\omega ^{\prime })=\frac{A(\omega ,\omega ^{\prime })}{\sqrt{%
\omega \omega ^{\prime }}},\beta (\omega ,\omega ^{\prime })=\frac{B(\omega
,\omega ^{\prime })}{\sqrt{\omega \omega ^{\prime }}}  \label{e106}
\end{equation}

The quantity $B(\omega ,\omega ^{\prime })$ is read off (\ref{bb9}) (and $%
A(\omega ,\omega ^{\prime })$ from the corresponding expression for $\alpha
(\omega ,\omega ^{\prime })$). From the expressions for the Bogolubov
coefficients one can immediately deduce that 
\begin{equation}
B^{*}(\omega ,\omega ^{\prime })=A(-\omega ,\omega ^{\prime }),A^{*}(\omega
,\omega ^{\prime })=B(-\omega ,\omega ^{\prime })  \label{e107}
\end{equation}

\noindent Observe that identities (\ref{e92a}), (\ref{92}) are a result of
the completeness of the basis {\it out }and {\it in }wavefunctions
respectively.

\section{A trajectory accelerating ad infinitum and the black body spectrum}

It was mentioned in the Introduction that one major mathematical difference
between type (i) and type (ii) trajectories lies in the behaviour of the $%
\beta \left( \omega ,\omega ^{\prime }\right) $ amplitude in the $\omega
^{\prime }\rightarrow \infty $ limit. To see this let us consider the
integral 
\begin{equation}
I\equiv \int_{0}^{\infty }dze^{i\omega ^{\prime }z-az}e^{-i\omega f(z)}
\label{t1}
\end{equation}

\noindent ($a$ being a convergence factor), which is one of the integrals
appearing in the amplitude (\ref{bb9}) (the other integrals are handled in a
similar way). To examine the asymptotic behaviour of (\ref{t1}) I split the
integral $\int_{0}^{\infty }$ to $\int_{0}^{r}+\int_{r}^{\infty }$ and apply
simple integration by parts {\it twice} to each one of them (see Bender and
Orszag \cite{BO}, p. 278). The endpoint contributions from infinity vanish
due to the convergence factor. The endpoint contributions at $z=r$ cancel
out in pairs: $f_{0}(r)$ cancels with $f_{acc}(r)$ (both accompanied by a
factor $1/\omega ^{\prime }$) and $f_{0}^{\prime }(r)$ cancels with $%
f_{acc}^{\prime }(r)$ (both accompanied by a factor $\left( \omega ^{\prime
}\right) ^{-2}$). These cancellations are hardly fortuitous. The first
cancellation reflects the fact that the trajectory itself $u=f(v)$ is
continuous, and the second reflects the continuity of the velocity at {\it P}%
. Similarly endpoint contributions at $z=0$ cancel with corresponding terms
originating from a large $\omega ^{\prime }$ expansion of the second
(trajectory independent) term in (\ref{bb9}) on the same continuity grounds
as above. Thus integral $I$ goes as $\left( \omega ^{\prime }\right) ^{-3}$
and observing the prefactors in (\ref{bb9}) we deduce that its contribution
to $\beta (\omega ,\omega ^{\prime })$ goes as $\left( \omega ^{\prime
}\right) ^{-5/2}$. The same behaviour is obtained for the other terms making
up $\beta (\omega ,\omega ^{\prime }).$

In the case of a type (ii) trajectory the amplitude is again given by (\ref
{bb9}) provided we replace the upper limit of the first integration by $\ln
2 $. It consists of the $\beta _{I}$ term given by (\ref{e20}) (again with $%
\infty $ replaced by $\ln 2$) and of the $\beta _{III}$ term given by (\ref
{e103}) (the $\beta _{II}$ term relates to the motion of the mirror after
point {\it P }and so it simply does not appear in a type (ii) trajectory).
The cancellation mechanism mentioned above does not exist now, and
integration by parts is of no help since the resulting integral diverges. I
thus adopt a different approach and transform the $\beta _{I}$ term to 
\begin{equation}
\beta _{I}(\omega ,\omega ^{\prime })=-\frac{2^{i\left( \omega -\omega
^{\prime }\right) }}{2\pi }\sqrt{\frac{\omega ^{\prime }}{\omega }}%
\int_{0}^{\ln 2}d\rho e^{i\omega ^{\prime }\rho }\left( 1-e^{-\rho }\right)
^{i\omega }  \label{bb24}
\end{equation}

To obtain the asymptotic behaviour of the above integral I employ the
standard technique of deforming the integration path to a contour in the
complex plane (see \cite{BO} chapter 6; see also Morse and Feshbach \cite{MF}%
, p. 610 where a very similar contour is used in the study of the asymptotic
expansion of the confluent hypergeometric). The deformed contour runs from 0
up the imaginary axis till $iT$ (I eventually take $T\rightarrow \infty $),
then parallel to the real axis from $iT$ to $iT+\ln 2$, and then down again
parallel to the imaginary axis from $iT+\ln 2$ to $\ln 2$. The contribution
of the segment parallel to the real axis vanishes exponentially in the limit 
$T\rightarrow \infty $. One is thus led to two integrals of the form 
\[
\int_{0}^{\infty }dse^{-\omega ^{\prime }s}\times (...) 
\]

\noindent Hence the $\omega ^{\prime }\rightarrow \infty $ limit is
equivalent to the $s\rightarrow 0$ limit. Eventually in the large $\omega
^{\prime }$ limit 
\begin{equation}
\beta _{I}(\omega ,\omega ^{\prime })\simeq -i\frac{2^{i\left( \omega
-\omega ^{\prime }\right) }}{2\pi \sqrt{\omega \omega ^{\prime }}}\left(
\omega ^{\prime }\right) ^{-i\omega }e^{-\pi \omega /2}\Gamma \left(
1+i\omega \right) +\frac{i}{2\pi \sqrt{\omega \omega ^{\prime }}}
\label{bb16}
\end{equation}

\noindent The second term in (\ref{bb16}) exactly cancels with the
asymptotic form of $\beta _{III}$ given by (\ref{e103}), the end result
being 
\begin{equation}
\beta (\omega ,\omega ^{\prime })\simeq -i\frac{2^{i\left( \omega -\omega
^{\prime }\right) }}{2\pi \sqrt{\omega \omega ^{\prime }}}\left( \omega
^{\prime }\right) ^{-i\omega }e^{-\pi \omega /2}\Gamma \left( 1+i\omega
\right)  \label{bb17}
\end{equation}

\noindent By taking the modulus of (\ref{bb17}) and squaring we get the
black body result 
\begin{equation}
\left| \beta (\omega ,\omega ^{\prime })\right| ^{2}\simeq \frac{1}{2\pi
\omega ^{\prime }}\frac{1}{e^{2\pi \omega }-1}  \label{lei5}
\end{equation}

\section{Conclusions}

I will try to summarize the main conclusions that can be drawn from the
above treatment , and also correct some misconceptions in the literature.

(a) It was shown that in the case of a type (ii) trajectory the amplitude
square $\left| \beta (\omega ,\omega ^{\prime })\right| ^{2}$ behaves as $%
1/\omega ^{\prime }$ (hence the corresponding spectrum integral diverges
logarithmically and a cut-off is required). This fact has been known for
some time, and is usually stated in the form that ''large frequencies
dominate''. However this statement is also unfortunately coupled to the
incorrect one that the important contributions to the black body spectrum
arise from the asymptotic part of the trajectory (i.e. the part lying close
to the asymptote $v=\ln 2$). It is shown in the Appendix of \cite{AC2} that
if one expands in powers of distance from the asymptote, then the
contributions from the various terms are all of the same order of magnitude
and thus the expansion cannot be truncated. Statements to the effect that
short distances dominate have been made by Hawking \cite{HAW} in the context
of radiation from black holes; their applicability in the present different
problem of an accelerating mirror should be doubted. In the present case the
misleading assertion that short distances dominate may be in accordance with
one's classical instincts, in the sense that roughly speaking the small
amount of time that the mirror spends near the origin ought to have an
insignificant effect compared to the infinitely long time spent near the
asymptote. It does however go counter to quantum mechanical orthodoxy,
according to which one cannot make statements as to where and when photons
have been produced. It is certainly true that were it not for the
singularity on the $v$ asymptote the thermal spectrum would not arise, but
this does not provide any guidance as to the origin of the dominant
contribution to the amplitude. This remark can be illustrated by the crucial
role of the $\beta _{III}$ term (missing in \cite{DF}) which appears {\it as
if } it originates at $t=0$.

One further danger in throwing away contributions has to do with possible
violations of unitarity. As already pointed out after (\ref{e103}), the $%
\zeta $ functions featuring in $\beta _{II}$, $\beta _{III}$ lead to the
delta function $\delta \left( \omega -\omega ^{\prime }\right) $ when one
calculates the $\alpha \left( \omega ,\omega ^{\prime }\right) $ amplitudes.
These delta functions are essential in ensuring the validity of the
unitarity relations (\ref{e92a}), (\ref{92}).

(b) Fulling and Davies produce perfectly good arguments in favour of the
black body spectrum for the type (ii) trajectory based on the calculation of
matrix elements of {\it local }field quantities. In this talk I have adopted
a rather different approach (also tackled in \cite{FD}, \cite{DF}) based on
the calculation of the Bogolubov coefficients. These quantities are by
definition time-independent, and in this context the question as to where
and when the photons are produced simply does not arise (this remark goes in
parallel to (a) above). Similarly attempts to distinguish between
''transient'' and ''steady state'' radiation at the level of the $\alpha $
and $\beta $ amplitudes are bound to fail; the emphasis in the literature on
the importance of the asymptotic part of the trajectory has unfortunately
led to the use of such terminology (see e.g. \cite{dew}).

\noindent {\bf Acknowledgments}

I am indebted to Professors Gabriel Barton and Stephen A. Fulling for
discussions and/or correspondence, and to Professor Michael Bordag for
kindly providing the opportunity to give this talk.

\end{document}